# Electrolyte Bonding Engineering for Highly Uniform GeTe-based CBRAM and Parallel Hebbian Learning in Selector-free Hopfield Networks

Jiin Bang[1,2], Jingyeong Hwang[1,3], Unhyeon Kang[1,3], Seungmin Oh[1,4], Kyungmin Lee[1,5] Jaehyun Park[1,6], Younghyun Lee[1], Hyun Jae Jang[1], Seongsik Park[1], YeonJoo Jeong[1], Inho Kim[1], Jong Keuk Park[1], Suyoun Lee[1,2*]

[1]*Center for Semiconductor Technology, Korea Institute of Science and Technology, Seoul 02792, Korea*

[2]*Nanoscience and Technology, KIST School, University of Science and Technology, Seoul 02792, Korea*

[3]*Department of Materials Science and Engineering, Seoul National University, Seoul 08826, Korea*

[4]*Department of Physics and Astronomy, Seoul National University, Seoul 08826, Korea*

[5]*Department of Electrical Engineering, Korea University, Seoul 02841, Korea*

[6]*Department of Materials Science and Engineering, Korea University, Seoul 02841, Korea*

**Email: S. Lee (slee_eels@kist.re.kr)*

**Hopfield networks offer a hardware-friendly framework for energy-efficient associative memory, yet their practical realization in memristor crossbar arrays is critically hindered by device-to-device (D2D) variability, which prevents reliable parallel programming. Here, we address this bottleneck through systematic composition engineering of the Ge–Te solid electrolyte in conductive bridge random access memory (CBRAM) devices. By varying the Ge:Te ratio, we identify $Ge_{3.5}Te_1$ as an optimal electrolyte composition that suppresses stochastic resistance variation by approximately three orders of magnitude compared to GeSe-based devices. Raman spectroscopy reveals**

**that this dramatic improvement originates from a bonding network dominated by asymmetric-stretching $GeTe_4$ tetrahedral units, which form interconnected free-volume channels that confine and stabilize $Cu^+$ ion migration pathways. Leveraging this enhanced uniformity, we fabricate a selector-less 16×16 Cu/$Ge_{3.5}Te_1$ CBRAM crossbar array and demonstrate a 4×4 Hopfield associative network capable of learning and recalling binary pattern pairs via fully parallel programming using a half-selection scheme. Successful pattern recall is achieved for up to two stored associations despite the absence of selector elements, establishing a proof-of-concept for selector-free hardware implementations of associative memory. These results highlight the critical role of electrolyte bonding structure in determining memristor uniformity and provide a materials-driven pathway toward scalable, parallel neuromorphic computing systems.**

## 1. Introduction

Recent advances in machine learning and semiconductor technologies have accelerated the development of artificial intelligence (AI) systems. In particular, large language models (LLMs) based on the Transformer architecture[1]–such as generative pretrained transformers (GPT)–achieve strong performance through an attention mechanism that relies on vector–matrix multiplication (VMM) followed by nonlinear Softmax operations. By ranking attention scores, these models effectively identify relevant context and reconstruct sequences from incomplete inputs. However, these performance gains come at substantial computational and energy costs. The complexity of attention scales as $O(n^2 \cdot d)$ with sequence length $n$ and representation dimension $d$, and the expansion to billion-parameter models further exacerbates the burden of repeated matrix multiplications[2]. This demand is fundamentally mismatched with conventional von Neumann architectures, in which data movement between physically separated memory and processing units dominates energy consumption — a bottleneck commonly known as the memory wall[3]. These limitations have driven growing interest in alternative computing paradigms for energy-efficient AI hardware.

A compelling alternative model inherently suited to pattern completion is the Hopfield network (see Figure 1a)[4]. In this architecture, a fully connected recurrent network of computational units called neurons stores associative patterns in a weight matrix via a local learning rule derived from Hebb's rule[5]. As an energy-based model, the network dynamics naturally drive the system toward the stored pattern closest to a given input, enabling pattern completion through iterative state updates. This parallel structure and simple learning rule afford fast learning and relatively low computational complexity. A recent theoretical advance — the modern Hopfield network (MHN) — further overcomes the storage capacity limit of classical Hopfield networks (approximately $0.14N$ for $N$ neurons) by achieving exponentially growing capacity with respect to $N$[6, 7]. Crucially, the use of the Softmax function as the MHN's activation function renders its mathematical formulation identical to that of the Transformer's attention layer. This structural equivalence has sparked renewed interest in Hopfield-type networks as promising, computationally efficient alternatives to the attention mechanism[8].

Memristor crossbar arrays offer a natural hardware platform for implementing Hopfield networks (Figure 1b)[9]. Memristors are two-terminal resistive switching devices whose nonvolatile, analog conductance-tuning characteristics enable direct representation of synaptic weights in hardware[10]. Accordingly, memristor crossbar arrays have been widely studied as VMM accelerators, offering substantial improvements in energy efficiency over conventional von Neumann architectures and graphics processing unit (GPU)-based systems[11, 12]. However, significant challenges remain in the hardware implementation of Hopfield networks — particularly in realizing parallel programming. The intrinsic stochasticity of resistive switching frequently introduces large device-to-device (D2D) and cycle-to-cycle (C2C) resistance variations. Although such variations can be mitigated through resistance-tuning approaches such as iterative write-and-verify methods in cell-by-cell programming, these techniques negate the computational benefits of parallel programming based on the Hebbian learning rule [13, 14]. Suppressing stochastic resistance variation is therefore essential for the practical implementation of memristor-based Hopfield networks. Several structural control strategies have been explored to this end, including the introduction of cone-like structures in the electrolyte to guide filament formation and reduce switching stochasticity[15]. Nevertheless, such approaches increase fabrication complexity and raise scalability concerns.

To address these challenges, we engineer the electrolyte material to establish a confined ion path within conductive bridge random access memory (CBRAM) devices integrated into a memristor crossbar array. Since CBRAM operation relies on the formation and rupture of a metallic filament within a solid electrolyte, spatially confining the ion path is expected to enhance switching reproducibility and D2D uniformity. We investigate $Ge_xTe_1$ and $Ge_{1.2}Se_1$ as candidate solid electrolytes, chosen for their well-documented high ionic conductivity[16, 17] and their rich diversity of local bonding structures – including corner-sharing and edge-sharing tetrahedral configurations ($GeTe_{4-n}Ge_n$, $n$ = 0-4) and trigonal pyramidal $GeTe_3$ units – which vary with composition[18, 19]. In an amorphous phase, this structural diversity is expected to increase the density of free volume, serving as a confined ion channel (see Figure 1c)[20, 21]. By systematically varying the Ge:Te ratio from 1.2:1 to 4.8:1, we characterize the switching behavior of the resulting CBRAM devices – including C2C and D2D resistance variability – and probe changes in local bonding structure via Raman spectroscopy. A strong correlation between device variability and local bonding configuration is identified, enabling the selection of an optimal composition for suppressing stochastic resistance variation. Finally, we fabricate a 4×4 cross-point CBRAM array and demonstrate an associative network capable of learning and recalling two distinct patterns via a fully parallel programming scheme based on a half-selection method.

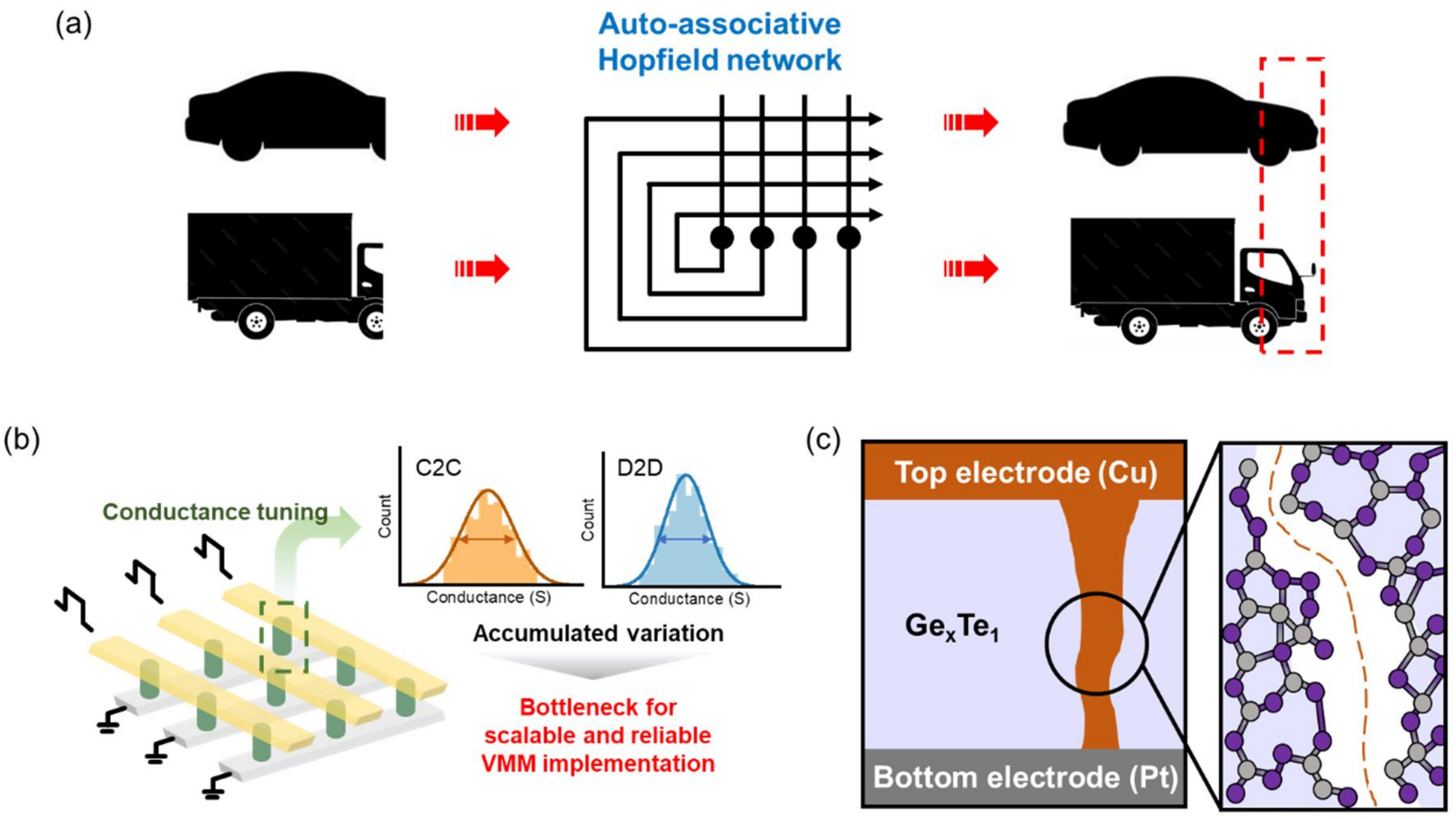

**Figure 1.** (a) Schematic illustration of auto-association using the Hopfield network. A partial (or corrupted) input pattern is presented to the network and reconstructed into the corresponding stored pattern. The computational units, i.e., neurons, are fully connected, and their states are updated iteratively through the recurrent network architecture. (b) Schematic illustration of a memristor crossbar array for learning and recall, in which synaptic weights are encoded as device conductance states. During conductance tuning via voltage pulses, cycle-to-cycle (C2C) and device-to-device (D2D) variations cause deviations from the target conductance values. The accumulated variation across the array degrades the accuracy of vector-matrix multiplication (VMM) operation and serves as a key bottleneck for scalable and reliable computation. (c) Schematic illustration of conductive filament formation in a $Cu/Ge_xTe_1/Pt$ CBRAM device. In the $Ge_xTe_1$ solid electrolyte, the local bonding structure gives rise to continuous free-volume pathways. These free-volume regions serve as confined ion channels that guide $Cu^+$ migration under an electric field, thereby localizing filament growth and enabling reproducible filament formation with improved switching uniformity.

## 2. Results and discussions

### 2.1. Composition-dependence of uniformity of CBRAM devices composed of $Ge_xTe_1$ and $Ge_{1.2}Se_1$

The CBRAM devices are fabricated in a pore-type structure to confine the current path, with pore diameters ranging from 200 to 500 nm ($d_{pore}$ = 200–500 nm), as shown in the cross-sectional transmission electron microscopy (TEM) image in Figure 2a. The CBRAM stack follows a Pt/$Ge_xTe_1$/Cu configuration, enabling systematic investigation of how solid electrolyte composition influences stochastic variations in device characteristics. $Ge_{1.2}Se_1$-based CBRAM is also investigated as a control sample. Film compositions are determined by Auger electron spectroscopy (AES), yielding four compositions – $Ge_{2.1}Te_1$, $Ge_{3.5}Te_1$, $Ge_{4.8}Te_1$, and $Ge_{1.2}Se_1$ – whose detailed depth profiles are provided in Supplementary Information Figure S1.

Figure 2b presents representative current–voltage (*I–V*) characteristics of CBRAM devices with $d_{pore}$ = 300 nm, measured under voltage sweep conditions and compliance current settings selected to yield the most stable switching behavior. Figure 2c summarizes the statistical distributions of $V_{\text{set}}$, $R_{\text{LRS}}$, and $R_{\text{HRS}}$ for devices employing each of the solid electrolytes, along with their corresponding mean values. As shown in Figure 2b and 2c, increasing Ge content in $Ge_xTe_1$ progressively reduces variability in $V_{\text{set}}$, $R_{\text{LRS}}$, and $R_{\text{HRS}}$. Most notably, D2D variation is dramatically suppressed at $x = 3.5$ and $x = 4.8$ in $Ge_xTe_1$, whereas substituting Te with Se ($Ge_{1.2}Se_1$) yields only marginal improvement in D2D uniformity. To quantitatively compare stochastic variation across devices, we define a parameter $\alpha = \prod_i (\sigma_i/\mu_i)$, where $\sigma_i$ and $\mu_i$ represent the standard deviation and the mean of each characteristic parameter ($i =$ $V_{\mathrm{set}}$, $R_{\mathrm{LRS}}$, and $R_{\mathrm{HRS}}$) extracted from ten devices (see Figure 2d). This analysis reveals that stochastic variation in $Ge_{3.5}Te_1$ is suppressed by approximately three orders of magnitude relative to $Ge_{1.2}Se_1$. In the following section, we demonstrate that this striking difference in $\alpha$ is intimately linked to distinctions in the local bonding structure of the solid electrolyte films.

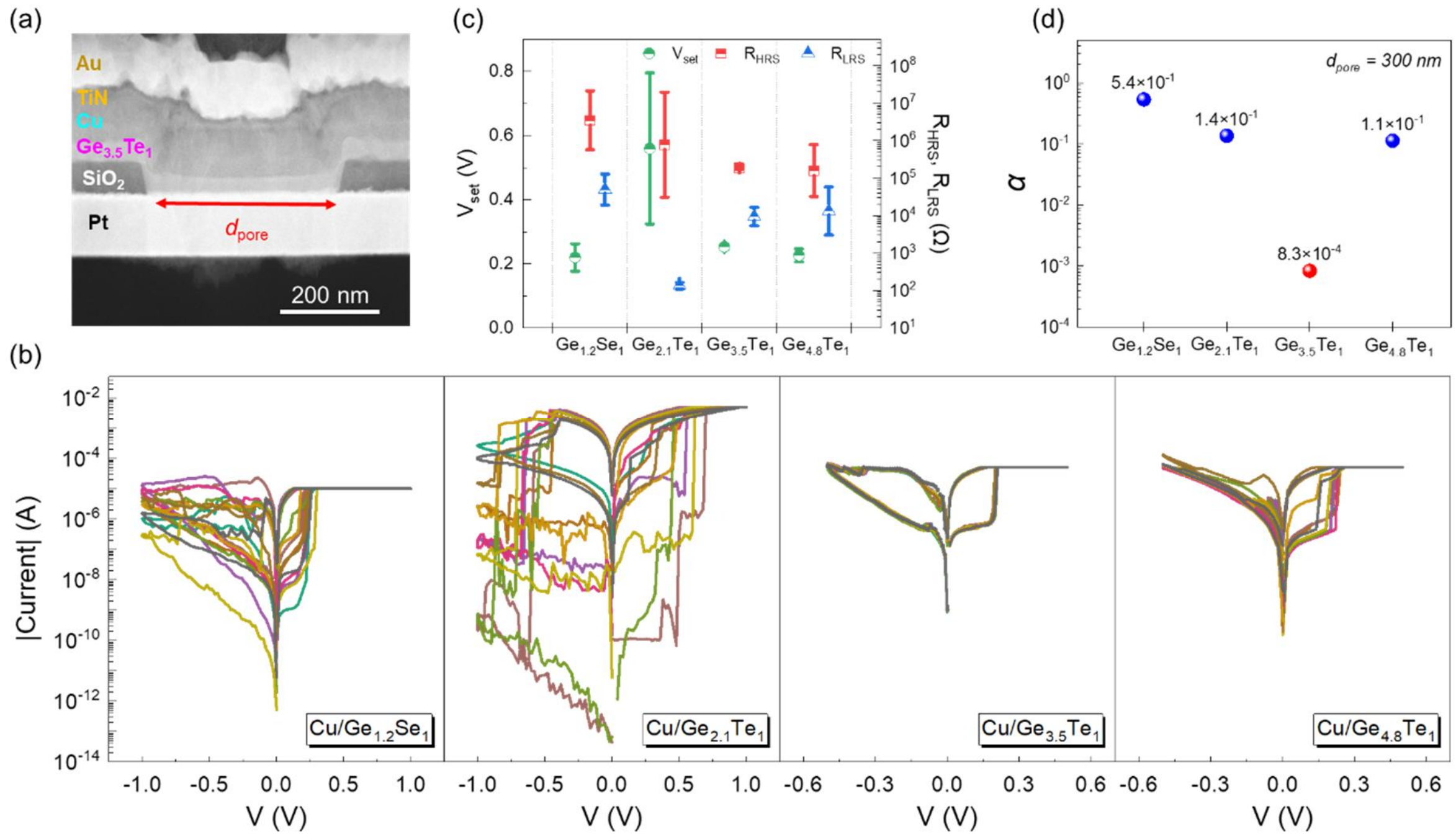

**Figure 2.** (a) Cross-sectional transmission electron microscopy (TEM) image of the fabricated CBRAM device. (b) Representative *I-V* characteristics of CBRAM devices with different solid electrolytes. The compliance current is set to a value that yields the most stable switching characteristics. Compared with $Ge_{1.2}Se_1$, $Ge_{2.1}Te_1$, and $Ge_{4.8}Te_1$, $Ge_{3.5}Te_1$ devices exhibit the most uniform switching behavior, indicating improved D2D uniformity. (c) Comparison of the $V_{\mathrm{set}}$, $R_{\mathrm{LRS}}$, and $R_{\mathrm{HRS}}$ with different solid electrolytes: $Ge_{1.2}Se_1$, $Ge_{2.1}Te_1$, $Ge_{3.5}Te_1$, and $Ge_{4.8}Te_1$. (d) Comparison of the variability in device characteristics, expressed as $\alpha = \prod_i(\sigma_i/\mu_i)$ ($i = V_{\mathrm{set}}$, $R_{\mathrm{LRS}}$, and $R_{\mathrm{HRS}}$) where $\sigma_i$ and $\mu_i$ represent the standard deviation and the mean of the characteristic values obtained from ten devices, respectively.

**2.2. Raman spectra: local bonding structures in $Ge_xTe_1$**

To clarify the relationship between observed device variability and the local bonding structure of the electrolyte material, we performed Raman spectroscopy analysis. Figure 3a shows Raman spectra for $Ge_{2.1}Te_1$, $Ge_{3.5}Te_1$, $Ge_{4.8}Te_1$, and $Ge_{1.2}Se_1$. The spectra exhibit markedly different features depending on composition, and the observed Raman peaks are assigned based on prior studies of bonding structures in Ge–Te and Ge–Se systems[18-22].

Ge atoms generally have a coordination number of four, while chalcogen atoms have a coordination number of two, forming tetrahedral units with a bond angle of 109.5° (Figure 3b). The Raman peaks associated with these tetrahedral units appear at 195 $cm^{-1}$ for $Ge_{1.2}Se_1$ and at 153 and 215 $cm^{-1}$ for $Ge_{3.5}Te_1$, respectively[18, 23]. In $Ge_{1.2}Se_1$, additional peaks at 113, 148, 167, and 280 $cm^{-1}$ reveal the coexistence of multiple bonding structures beyond simple tetrahedral units. Specifically, the peaks at 113 and 148 $cm^{-1}$ correspond to tetrahedral units within the orthorhombic phase of crystalline $Ge_{1.2}Se_1$. The 167 $cm^{-1}$ peak arises from ethane-like Ge–Ge homopolar units, in which each Ge atom is bonded to three Se atoms. The 280 $cm^{-1}$ peak corresponds to Se–Se chain structures formed by successive Se homopolar bonds [24]. Taken together, these features indicate that $Ge_{1.2}Se_1$ contains a complex mixture of tetrahedral, orthorhombic, Ge–Ge, and Se–Se bonding configurations.

In contrast, $Ge_{2.1}Te_1$ exhibits a simpler spectrum dominated by two main peaks near 122 and 160 $cm^{-1}$, with satellite peaks near 108, 175, and 220 $cm^{-1}$. The peaks near 122 and 175 $cm^{-1}$ are assigned to $Ge_1Te_2Ge_2$ and $Ge_1Te_3Ge_1$ tetrahedra, respectively, in which Ge atoms partially substitute Te within the tetrahedral framework (Figure 3b)[18]. The peaks near 175–180 $cm^{-1}$ are attributed to the symmetric stretching ($A_1$-like) mode of $GeTe_4$ tetrahedra, while the peak near 225 $cm^{-1}$ corresponds to the asymmetric stretching ($F_2$-like) mode of GeTe[18]. In $Ge_{3.5}Te_1$, the peak near 108 $cm^{-1}$ is strongly suppressed while the spectral weight near 220 $cm^{-1}$ increases markedly, indicating that the $F_2$-like mode of GeTe becomes the dominant structural motif. With further increases in Ge content, $Ge_{4.8}Te_1$ exhibits a new dominant peak near 273 $cm^{-1}$ attributed to amorphous Ge clusters[22], while the characteristic peaks of $Ge_{2.1}Te_1$ and $Ge_{3.5}Te_1$ become negligible.

These differences in bonding structure are expected to influence $Cu^+$ migration strongly and, consequently, the repeated formation and rupture of conductive filaments. In amorphous $Ge_{3.5}Te_1$, the dominant tetrahedrally coordinated Ge–Te motifs form a relatively open and flexible amorphous network with a high density of interconnected free-volume regions. These extended structural voids act as percolative ion-transport pathways, lowering the activation energy barrier for $Cu^+$ migration under an applied field and thereby facilitating long-range ion transport and the formation of continuous conductive bridges (Figure 3c)[25]. In contrast, amorphous $Ge_{1.2}Se_1$ and $Ge_{4.8}Te_1$ – enriched with Se–Se and Ge–Ge homopolar bonds, respectively – form more rigid, densely packed subnetworks characterized by fewer continuous void channels and more spatially isolated nanovoids (Figure 3d). $Ge_{2.1}Te_1$ is similarly expected to exhibit a relatively rigid network, as its local bonding structure is dominated by $Ge_1Te_2Ge_2$, $Ge_1Te_3Ge_1$, and $A_1$-like $GeTe_4$ motifs. Consequently, the reduced connectivity of free-volume regions and the lower fraction of heteropolar Ge–Te coordination collectively diminish the number of low-energy interstitial sites available for $Cu^+$ conduction, localizing ion-transport pathways and increasing both the forming voltage and variability of conductive filament formation.

In $Ge_{3.5}Te_1$, by contrast, the simplified bonding network dominated by $GeTe_4$ tetrahedra – together with the presence of Ge–Ge ring structures – confines $Cu^+$ migration to well-defined pathways, enabling uniform and repeatable filament formation and rupture. These findings are consistent with the significantly enhanced switching uniformity observed in $Ge_{3.5}Te_1$-based CBRAM devices relative to GeSe-based devices.

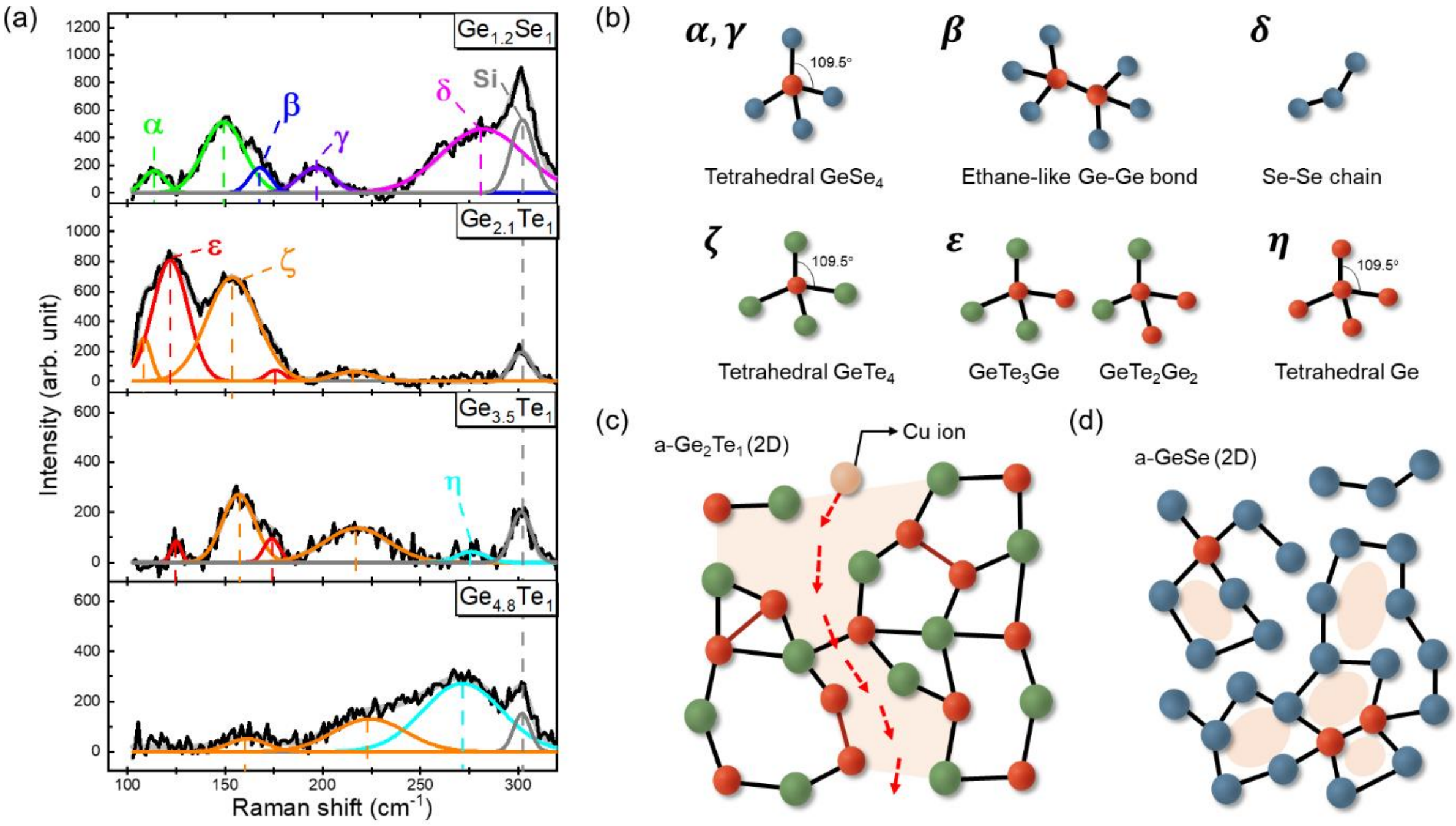

**Figure 3.** (a) Raman spectra of $Ge_{1.2}Se_1$ and $Ge_xTe_1$ ($x$ = 2.1, 3.5, and 4.8) films, with individual peaks resolved by multi-peak Gaussian fitting. The symbols corresponding to each peak are defined in (b). (b) Bonding structures identified in $Ge_xTe_1$ and $Ge_{1.2}Se_1$ thin films. (c) Schematic illustration of ring networks in amorphous $Ge_{3.5}Te_1$. (d) Schematic illustration of isolated free volumes in amorphous $Ge_{1.2}Se_1$. Green, red, blue, and pink dots denote Te, Se, Ge, and Cu atoms, respectively.

### 2.3. Parallel programming of the 4×4 Hopfield associative network composed of $Pt/Ge_{3.5}Te_1/Cu$ memristors

Based on the results presented in Figure 2, we selected a $Ge_{3.5}Te_1$-based CBRAM device to implement the Hopfield network. To further optimize uniformity of cells in the array, we investigated the effect of pore size ($d_{pore}$; see Figure 2a) and found that uniformity improves monotonically with decreasing $d_{pore}$ (see Supplementary Information Figure S2). A 16×16 crossbar array of $Pt/Ge_{3.5}Te_1/Cu$ memristor devices with a 200 nm pore size was subsequently fabricated (Figure 4a). To characterize the analog programmability of individual CBRAM cells, we first examined potentiation and depression (P/D) behavior as a function of pulse amplitude and width. We find that several tens of distinguishable analog conductance levels can be reliably programmed within the range of 0.75–3.5 mS by applying pulses of (200 μs, 0.55 V) and (200 μs, −0.5 V) for SET and RESET operations, respectively (see Supplementary Information Figure S4).

Using this CBRAM array, we implemented pattern-associative learning within a classical (Pavlovian) conditioning framework[26] by pairing a conditioned stimulus (CS) with an unconditioned stimulus (UCS). Synaptic weights are updated according to a Hebbian learning rule applied to each paired stimulus pattern (see Supplementary Information Note 1 for details). To enable parallel updating of synaptic weights, the CBRAM crossbar array adopts a 0T1R (zero-transistor, one-resistor/memristor) architecture in conjunction with a half-selection scheme[27], in which half the programming voltage ($V_{prg}/2$) is applied to the top electrode (TE) and $-V_{prg}/2$ to the bottom electrode (BE) of each selected cell (see Figure 4b). Due to sneak path currents arising from the absence of a selector device and the self-rectifying characteristics of the CBRAM cell, the required programming voltage was found to be $V_{prg}$ = 2 V − substantially higher than that needed for discrete cells (see Supplementary Information Figure S3 and S4). The effective array block size for parallel programming was therefore limited to 4×4.

Following initialization (see Experimental Methods), the CBRAM array was programmed with UCS-CS pattern pairs by applying the corresponding voltages to the BE and TE lines, respectively, within the half-selection scheme. Pattern-associative learning was performed sequentially for three stimulus pairs: [(UCS), (CS)] = [(1100), (1010)], [(0011), (0101)], and [(1001), (0110)]. Figure 4c shows the evolution of the conductance map across sequential training steps. After applying the first pattern pair, the conductance of the selected cells ($G_{11}$, $G_{12}$, $G_{32}$, and $G_{33}$) increased by an average of 38 μS, while unselected cells exhibited negligible change. After training the second pattern pair [(0011), (0101)], two of the four selected cells ($G_{23}$ and $G_{44}$) increased in conductance by an average of 84 μS, whereas the remaining two ($G_{24}$ and $G_{43}$) showed little response. Finally, after training the third pattern pair [(1001), (0110)], only one cell exhibited conductance increase. These results indicate that parallel programming of a 0T1R CBRAM array is susceptible to occasional errors, attributable primarily to three factors: (i) sneak path currents inherent to the selector-less architecture, which can induce unintended switching in half-selected cells; (ii) the stochastic nature of filament formation in CBRAM devices; and (iii) the absence of reliable state verification during parallel programming, which allows incorrectly programmed cells to go undetected and uncorrected.

Despite the presence of programming errors, successful recall of the UCS pattern from its associated CS pattern is achieved, owing to the energy-based memory principle and the integrate-and-fire (I&F) neuron dynamics of the Hopfield network. During the recall stage, the CS pattern is applied to the TE rows of the memristor array, with logic '1' encoded as $V_{read}$ and logic '0' as 0 V, while all BE columns are held at ground. By Ohm's law and Kirchhoff's current law, the current ($I_{j,\ read}$) measured at the $j$-th BE column corresponds to the activation value obtained via VMM;

$$I_j = \sum_i G_{ij} V_i \qquad \dots (1)$$

where $G_{ij}$ is the conductance of the memristor at the intersection of the $i$-th TE row and the $j$-th BE column, and $V_i$ is the applied input voltage. The resulting current is compared against a threshold current $I_{th}$ and binarized to logic '1' or '0'. Successful recall is confirmed when the binarized output matches the target UCS.

Figure 4d shows the evolution of $I_{j,read}$ as associative learning accumulates across training steps. With $I_{th}$ = 4 μA, recall remains successful through the first two pattern pairs. Notably, despite the presence of conductance errors following training of the second set, recall is still performed correctly, demonstrating that the Hopfield network's energy-based dynamics confer a degree of tolerance to minor programming defects in the CBRAM array. After training the third pattern pair, however, recall fails for both the second and third sets. As evidenced by the conductance map in the final state (rightmost panel of Figure 4c), this failure is attributed to the limited storage capacity of the Hopfield network implemented on the 4×4 CBRAM array.

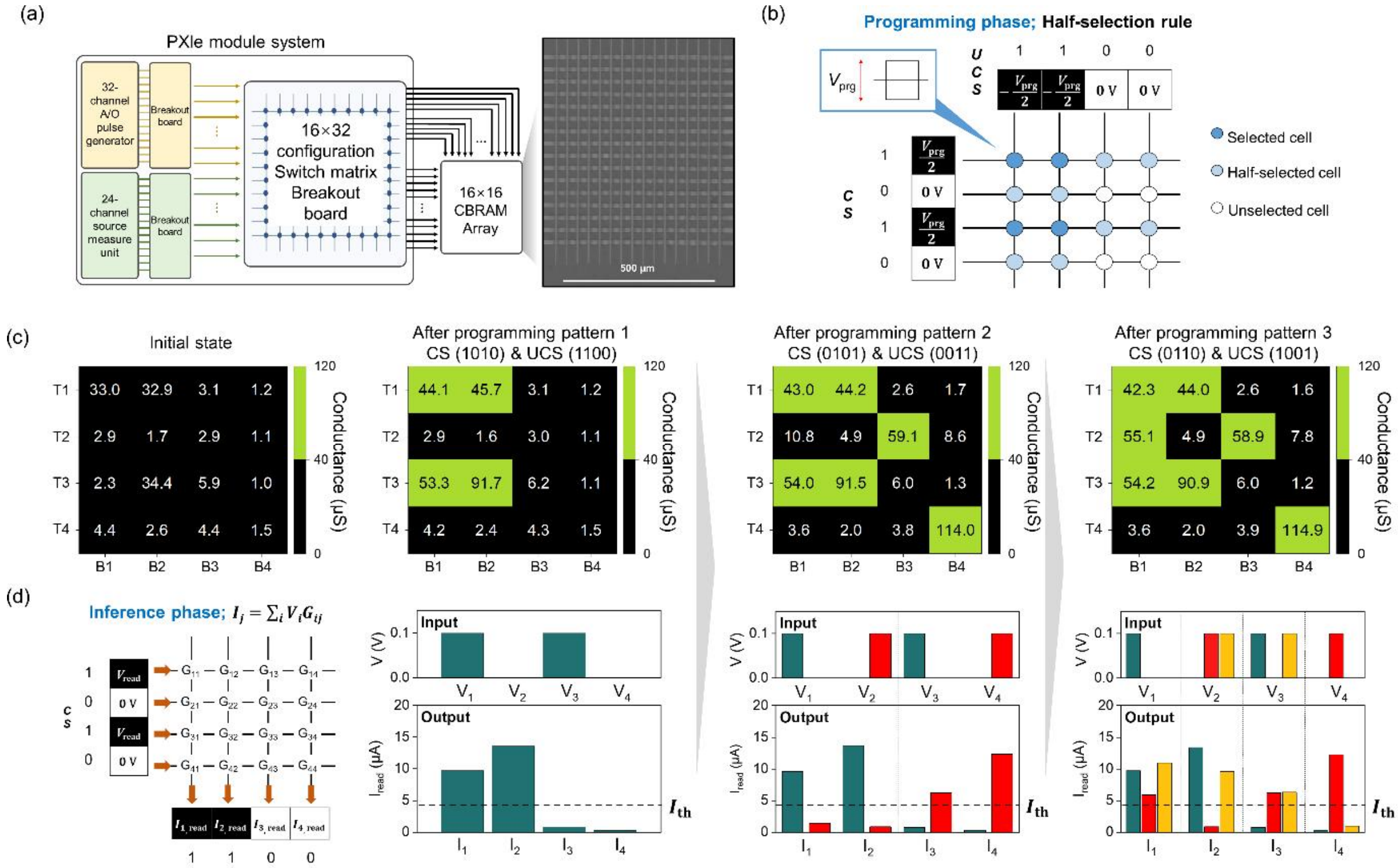

**Figure 4.** (a) Measurement system for parallel programming and inference of the 16×16 Cu/$Ge_{3.5}Te_1$ CBRAM cross-point array, along with a top-view scanning electron microscopy (SEM) image of the fabricated array. (b) Schematic of a half-selection scheme for parallel programming for hardware implementation of classical conditioning-based pattern association. (c) Conductance map in a 4×4 array following sequential learning of CS-UCS pattern pairs 1, 2, and 3 ($V_{\mathrm{prg}}$ = 2 V, 100 μs). (d) Schematic and experimental results of associative inference performed by the Hopfield network after sequential learning of three CS-UCS pairs ($V_{\mathrm{read}}$ = 0.1 V, $I_{\mathrm{th}}$ = 4 μA).

## 3. Conclusion

Hopfield networks, driven by the collective dynamics of coupled neurons, provide a compelling framework for energy-efficient associative memory and pattern completion. A key challenge in their hardware realization lies in achieving parallel programming of synaptic weights, as device variability and sneak path currents in memristor arrays typically necessitate sequential write-and-verify schemes that fundamentally undermine the benefits of parallelism.

In this work, we addressed this challenge through material-level engineering of the solid electrolyte in CBRAM devices. By systematically tuning the Ge–Te composition, we identified $Ge_{3.5}Te_1$ as an optimal electrolyte matrix that significantly enhances device uniformity. Raman spectroscopy revealed that this improvement is intimately associated with the dominance of tetrahedral GeTe units, which form a network of interconnected free-volume pathways that facilitate stable $Cu^+$ ion migration and reproducible conductive filament formation.

Leveraging this enhanced uniformity, we experimentally demonstrate associative learning in a 4×4 selector-less CBRAM crossbar array via a fully parallel, one-step Hebbian update scheme. Unlike conventional approaches that rely on iterative or cell-by-cell programming, the proposed method enables simultaneous weight updates across the entire array, directly realizing the core learning mechanism of the Hopfield network in hardware. Despite the programming errors arising from sneak path currents and stochastic filament formation, successful pattern recall is achieved − highlighting the inherent robustness of energy-based network dynamics against device-level imperfections.

While practical large-scale implementations are generally expected to employ 1S1R architectures incorporating dedicated selector devices, our results demonstrate that parallel programming and associative learning are achievable even under minimal hardware constraints.

This work thus establishes a proof-of-principle for one-step Hebbian learning in selector-less memristor arrays and offers new insight into the design of simplified, scalable, and energy-efficient neuromorphic computing systems.

## Experimental Methods

### Device fabrication

Pore-type CBRAM devices were fabricated on p-type Si (100) wafers with a 5000 Å $SiO_2$ layer. After substrate cleaning, 500 µm × 500 µm TiN/Ti/Pt BE were formed by reactive RF sputtering and e-beam evaporation, followed by lift-off patterning. A 50-nm $SiO_2$ insulating layer was then deposited by Plasma-enhanced chemical vapor deposition (PECVD), and nanoscale pores with diameters of 100-500 nm were defined by electron-beam lithography and Reactive ion etching (RIE). After top electrode patterning, $Ge_{1.2}Se_1$ or $Ge_xTe_1$ solid electrolyte layers were deposited by RF magnetron sputtering. Subsequently, a 30-nm Cu ion-supplying layer and a 50-nm TiN passivation layer were in situ deposited. The TE area was 500 µm × 500 µm, with a 100 µm × 100 µm overlap region containing the switching region. Finally, a Ti/Au capping layer was deposited by e-beam evaporation to suppress electrode oxidation.

### Characterization

Cross-sectional images of fabricated devices were obtained by transmission electron microscopy using focused ion beam (FIB-TEM) analysis. The Helios G4 HX model was used for FIB analysis (accelerating voltage 5 kV, source current 0.2 nA, ETD detector), and the Talos model was used for TEM image acquisition and EDS mapping. The composition of solid electrolyte films was analyzed by Auger electron spectroscopy (AES) depth profiling using a PHI 710 system (ULVAC-PHI, Inc.) equipped with a cylindrical mirror analyzer (CMA) (See Supplementary Information Figure S1). The electron beam energy was 5 kV; 50 nm-thick films were deposited on p-type Si substrates with a 5000 Å $SiO_2$ layer for AES measurements.

The characteristic *I-V* curves of the CBRAM device shown in Figure 2b and P/D characteristic of the CBRAM device shown in Supplementary Information Figure S4 have been measured using a semiconductor characterization system (TEKTRONIX Keithley 4200A). Raman spectroscopy was performed using a Renishaw inVia Raman microscope with a 785 nm excitation laser. The laser power was set to 5%, and the exposure time was 10 s. All measurements were performed at room temperature under ambient conditions.

**Pattern-association programming**

CBRAM array programming and inference measurements were performed using a PXIe module system consisting of a 32-channel analog-output (A/O) pulse generator (PXIe-6138) and a 24-channel source measure unit (SMU) (PXIe-4163), connected via breakout boards interfaced with a 16×32 configured switch matrix (PXIe-2532B).

Before pattern-association programming, the array is initialized to construct the initial weight matrix in a column-by-column manner by applying $V_{reset}$ pulses (100 μs, incremental-step amplitude) to TE lines while BE lines are grounded. BE lines of the remaining columns are biased at the same $V_{reset}$ as an inhibition bias. This initialization is conducted column-by-column with an ending condition of $G < 10$ μS, where $G$ is the conductance of a CBRAM cell.

## Author Information

**Corresponding author**

**Suyoun Lee –** *Center for Semiconductor Technology, Korea Institute of Science and Technology, Seoul 02792, Rep. of Korea; Nanoscience and Technology, Korea University of Science and Technology, Daejon 34136, Rep. of Korea;* https://orcid.org/0000-0002-5147-6821*; Email:* slee_eels@kist.re.kr

**Authors**

**Jiin Bang -** *Center for Semiconductor Technology, Korea Institute of Science and Technology, Seoul 02792, Rep. of Korea; Nanoscience and Technology, KIST school, Korea University of Science and Technology, Seoul 02792, Rep. of Korea*

**Jingyeong Hwang -** *Center for Semiconductor Technology, Korea Institute of Science and Technology, Seoul 02792, Rep. of Korea; Department of Materials Science and Engineering, Seoul National University, Seoul 08826, Korea*

**Unhyeon Kang -** *Center for Semiconductor Technology, Korea Institute of Science and Technology, Seoul 02792, Rep. of Korea; Department of Materials Science and Engineering, Seoul National University, Seoul 08826, Korea;*

**Seungmin Oh -** *Center for Semiconductor Technology, Korea Institute of Science and Technology, Seoul 02792, Rep. of Korea; Department of Physics and Astronomy, Seoul National University, Seoul 08826, Korea*

**Kyungmin Lee -** *Center for Semiconductor Technology, Korea Institute of Science and Technology, Seoul 02792, Rep. of Korea; Department of Electrical Engineering, Korea University, Seoul 02841, Korea*

**Jaehyun Park -** *Center for Semiconductor Technology, Korea Institute of Science and Technology, Seoul 02792, Rep. of Korea; Department of Materials Science and Engineering, Korea University, Seoul 02841, Korea*

**Younghyun Lee -** *Center for Semiconductor Technology, Korea Institute of Science and Technology, Seoul 02792*

**Hyun Jae Jang -** *Center for Semiconductor Technology, Korea Institute of Science and Technology, Seoul 02792*

**Seongsik Park -** *Center for Semiconductor Technology, Korea Institute of Science and Technology, Seoul 02792*
**YeonJoo Jeong -** *Center for Semiconductor Technology, Korea Institute of Science and Technology, Seoul 02792*
**Inho Kim -** *Center for Semiconductor Technology, Korea Institute of Science and Technology, Seoul 02792*
**Jong Keuk Park -** *Center for Semiconductor Technology, Korea Institute of Science and Technology, Seoul 02792*

**Author Contributions**

S.L. designed and conceived the experiments with help from J.B., J.H., and U.K. and J.B. fabricated CBRAM devices. J.B., J.P., I.K., and Y.J.J performed the characterization of the CBRAM devices. S. O., K.L., and Y.L. contributed to the measurement setup and designed parallel programming. J.B., H.J.J., S.P., and S.L. participated in the data analysis and interpretation. All authors discussed the data and participated in the writing of the manuscript.

**Notes**

The authors declare no competing financial interests.

## Acknowledgments

This work was supported by the Korea Institute of Science and Technology (KIST, Grant No. 26E0020) and by the National Research Council of Science & Technology (NST, Grant No. GTL24041-000) and National Research Foundation of Korea (NRF, Grant No. RS-2025-24533987) by the Korea government (MSIT).

## Supplementary Information for

# Electrolyte Bonding Engineering for Highly Uniform GeTe-based CBRAM and Parallel Hebbian Learning in Selector-free Hopfield Networks

Jiin Bang[1,2], Jingyeong Hwang[1,3], Unhyeon Kang[1,3], Seungmin Oh[1,4], Kyungmin Lee[1,5] Jaehyun Park[1,6], Younghyun Lee[1], Hyun Jae Jang[1], Seongsik Park[1], YeonJoo Jeong[1], Inho Kim[1], Jong Keuk Park[1], Suyoun Lee[1,2*]

[1]*Center for Semiconductor Technology, Korea Institute of Science and Technology, Seoul 02792, Korea*

[2]*Nanoscience and Technology, KIST School, University of Science and Technology, Seoul 02792, Korea*

[3]*Department of Materials Science and Engineering, Seoul National University, Seoul 08826, Korea*

[4]*Department of Physics and Astronomy, Seoul National University, Seoul 08826, Korea*

[5]*Department of Electrical Engineering, Korea University, Seoul 02841, Korea*

[6]*Department of Materials Science and Engineering, Korea University, Seoul 02841, Korea*

**Email: S. Lee (slee_eels@kist.re.kr)*

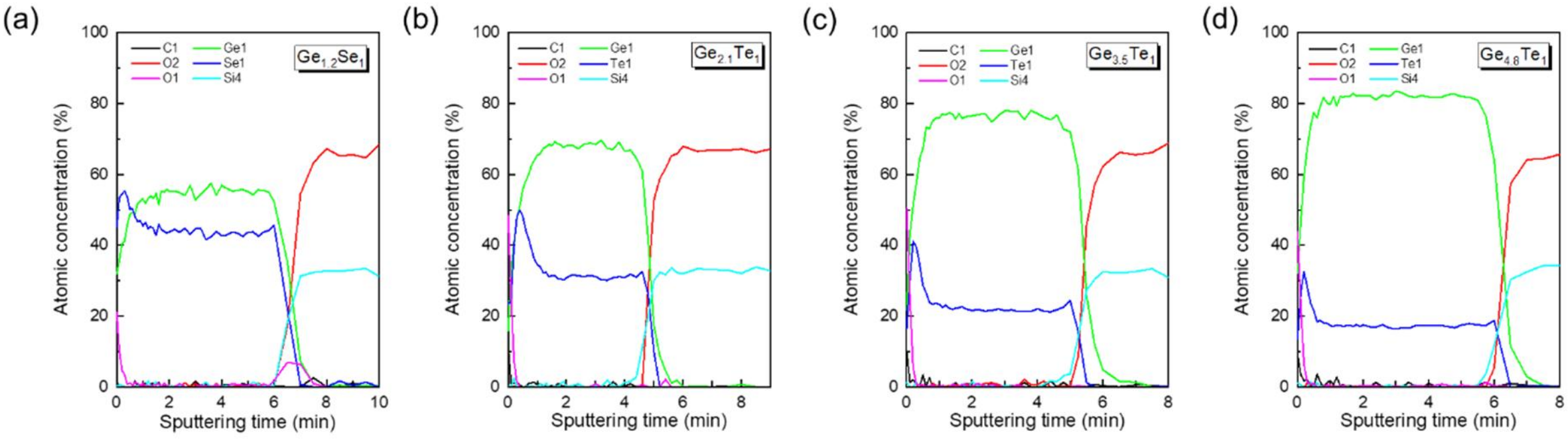

**Figure S1.** Auger electron spectroscopy (AES) analysis of the solid electrolyte thin films. The films were deposited by RF magnetron sputtering using GeSe, Ge, and GeTe targets. The measured compositions were (a) $Ge_{1.2}Se_1$ (b) $Ge_{2.1}Te_1$ (c) $Ge_{3.5}Te_1$ (d) $Ge_{4.8}Te_1$.

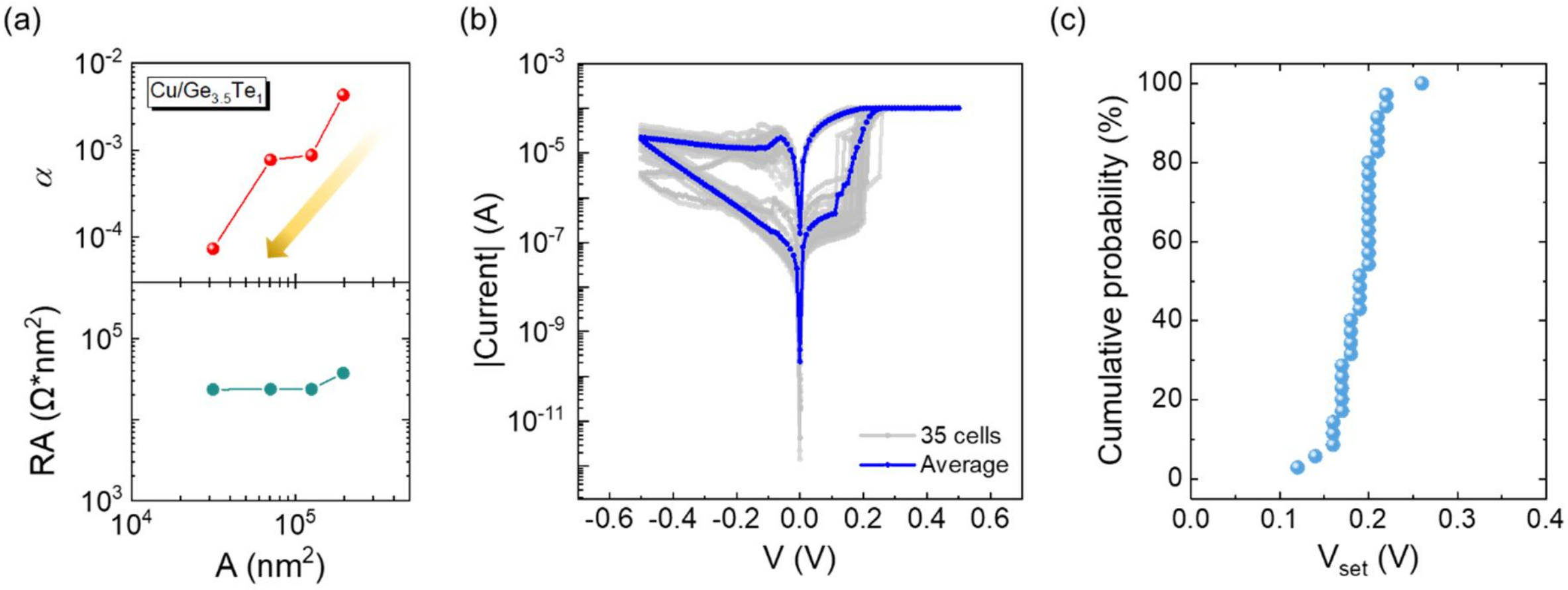

**Figure S2.** (a) Switching area dependence of the $\alpha$ of $Cu/Ge_{3.5}Te_1$ CBRAM devices. The switching area is given by $A = \pi(d/2)^2$, where $d$ is the pore diameter of 200, 300, 400 and 500 nm. For each pore size, five cells were used to extract the $\alpha$ values. (b) Representative I-V characteristics of 35 individual 200 nm $Cu/Ge_{3.5}Te_1$ CBRAM cells. (c) Cumulative probability plot of $V_{set}$ for the 35 cells, showing a median value of 0.2 V.

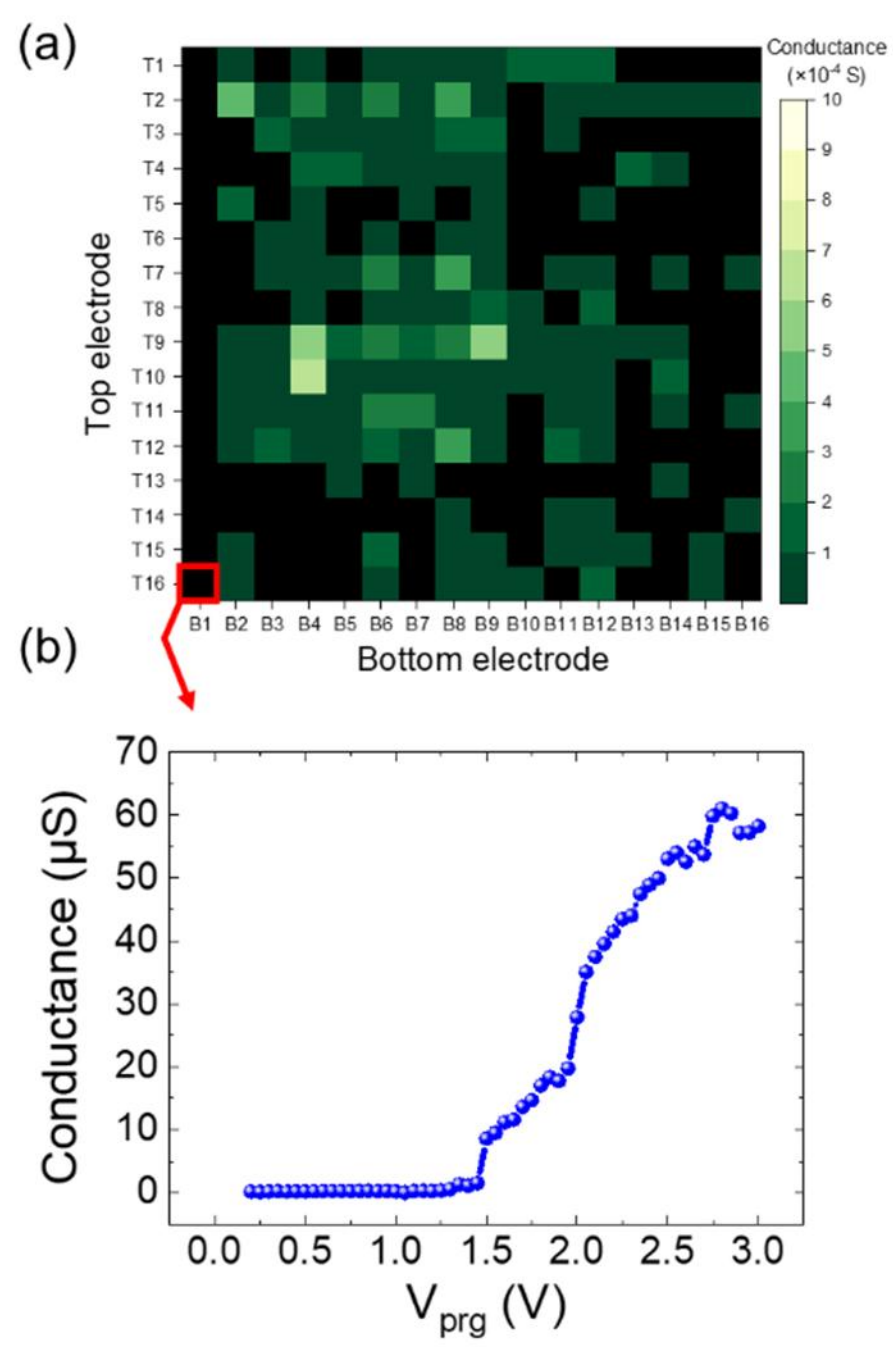

**Figure S3.** (a) Heat map of the conductance states in 16×16 Cu/$Ge_{3.5}Te_1$ CBRAM array with 200 nm pore size. (b) Conductance evolution of the T16B1 cell as a function of $V_{\text{prg}}$. The T16B1 cell was selected by biasing its TE and BE, while all other electrodes in the array were left floating. Incremental step pulse programming (ISPP) was performed by sweeping $V_{\text{prg}}$ from 0.3 to 3.0 V with a step size of 0.05 V and a pulse width of 100 μs. During measurement, $+1/2\,V_{\text{prg}}$ and $-1/2\,V_{\text{prg}}$ were applied to the TE and BE of the selected cell, respectively, and the conductance was read after each pulse to monitor the conductance evolution. The conductance exceeded 20 μS at $V_{\text{prg}}$ = 2 V, corresponding to a resistance below 50 kΩ, which was defined as the LRS. Therefore, $V_{\text{prg}}$ = 2 V was selected as the programming voltage for the array operation.

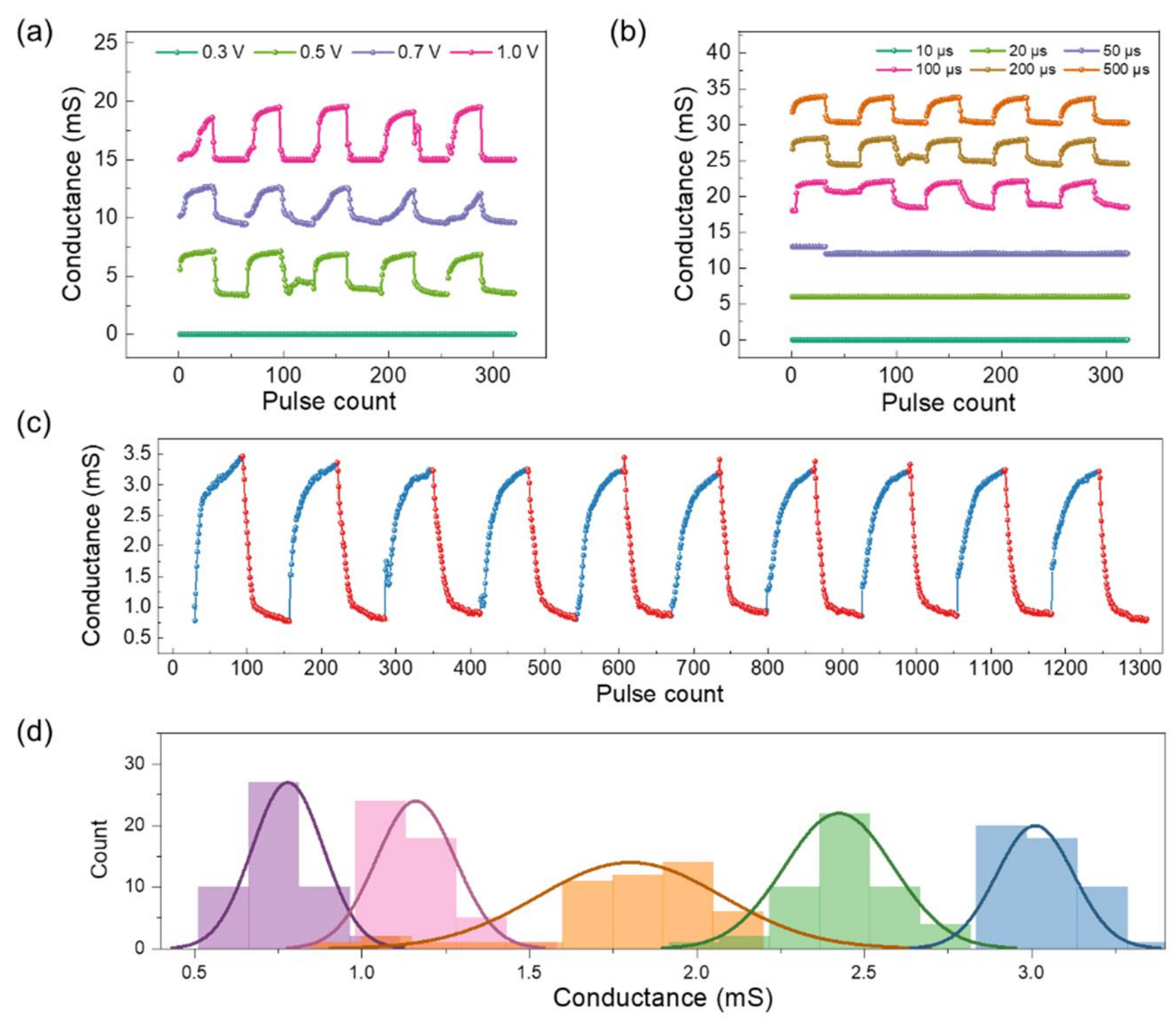

**Figure S4.** Potentiation/depression (P/D) characteristics of Cu/$Ge_{3.5}Te_1$ CBRAM devices with a 200-nm pore. Conductance was read at 0.1 V after each programming pulse. Each P/D cycle consisted of 64 potentiation pulses followed by 64 depression pulses. (a) Pulse-amplitude dependence of the P/D characteristics measured at a fixed pulse width of 200 μs. The pulse amplitude was varied from ±0.3 V to ±1.0 V, with positive pulses for potentiation and negative pulses used for depression. For clarity, the P/D curves measured at ±0.5 V, ±0.7 V, and ±1.0 V were vertically offset by 3, 9, and 15 mS, respectively. (b) Pulse-width dependence of the P/D characteristics measured at a fixed pulse amplitude of ±0.5 V. The pulse width was varied from 10 to 500 μs, and the P/D curves were vertically offset by 3, 9, 15, 21, and 27 mS, respectively. (c) Ten consecutive P/D cycles were measured under the optimized pulse conditions of +0.55 V for potentiation and -0.5 V for depression, with a pulse width of 200 μs, demonstrating stable and reproducible analog conductance modulation over repeated cycling. (d) Histogram of five programmed conductance levels obtained by pulse-based conductance modulation. The distributions were fitted with Gaussian functions to evaluate the separation and uniformity of the analog state. The mean conductance values were approximately 0.8, 1.2, 1.8, 2.4 and 3.0 mS, with standard deviations of 0.1, 0.12, 0.27, 0.15, and 0.11 mS, respectively.

**Supplementary Note 1. Pattern-associative learning**

Classical conditioning, originally established from Pavlov's studies on conditioned reflexes, is a form of associative learning in which a previously neutral stimulus becomes capable of eliciting a response after repeated pairings with a biologically meaningful stimulus.[1] Building on this principle, a pattern associator can be designed to store and recall paired patterns. In this work, we implement such an associative-learning process using the CBRAM array as a Hopfield associative network.[2] In detail, each CS-UCS training pattern pair is represented by two binary patterns of size *N*, denoted as $\mathbf{x}^{\mu} = (x_1^{\mu}, x_2^{\mu}, \cdots, x_N^{\mu})$ for the CS pattern and $\mathbf{y}^{\mu} = (y_1^{\mu}, y_2^{\mu}, \cdots, y_N^{\mu})$ for the corresponding UCS pattern, where *N* is the total number of neurons and $\mu \in \{1, 2, \cdots, p\}$ indexes the patterns. Here, each neuronal state is binary, taking the value 1 for a firing neuron and 0 otherwise. To learn the association between the CS and UCS patterns, the synaptic weights are updated according to a Hebbian learning rule between the two paired patterns. Starting from an initial zero matrix, $w_{ij}^{(0)} = 0$, the weights are cumulatively updated after each pattern learning, and the final synaptic weight matrix is expressed as

$$w_{ij} = \sum_{\mu} x_i^{\mu} y_j^{\mu}, \quad i \neq j \tag{1}$$

where the summation is taken over all CS-UCS pairs.

After learning, when only a CS pattern is applied as an input, the activation of the *j*-th output neuron is obtained by

$$h_j = \sum_i w_{ij} x_i \qquad (2)$$

and the output neuronal state is determined by thresholding the activation value:

$$y_i = \begin{cases} 1, & h_j \geq \theta_j \\ 0, & h_j < \theta_j \end{cases} \qquad (3)$$